\documentclass[twocolumn,10pt,showpacs,preprintnumbers,amssymb,aps, prl]{revtex4-1}
\usepackage{graphicx}
\usepackage{dcolumn}
\usepackage{bm}
\usepackage{array}
\usepackage{amstext}
\usepackage{latexsym,epsfig}
\newcommand{\be}{\begin{equation}}
\newcommand{\ee}{\end{equation}}
\newcommand{\ba}{\begin{eqnarray}}
\newcommand{\ea}{\end{eqnarray}}
\newcommand{\ban}{\begin{eqnarray*}}
\newcommand{\ean}{\end{eqnarray*}}
\def\a{\alpha}
\begin{document}

\title{Relation between two approaches of the colour hydrodynamics}
\vspace*{0.5cm}

\author{Abhishek~Basak and Jitesh~R.~Bhatt}
\email{jeet@prl.res.in}
\affiliation{Theoretical Physics Division, Physical Research Laboratory, Ahmedabad-380009, India}

\author{Predhiman~K.~Kaw}
\affiliation{Institute for Plasma Research, Bhat, Gandhinagar, India}
\begin{abstract}
It is argued that the short time scale phenomena can be studied
within the framework of hydrodynamics in the quark-gluon plasma.
There are two different versions of the hydrodynamic-like equations
in the literature. In this work we discuss the possible relationship
between these versions. In particular we show that if the colour charges
associated with the velocity and density matrices in the matrix version
of hydrodynamics are same then both the versions of the hydrodynamics
become identical.
\end{abstract}
\maketitle
Due to its simplicity over the other descriptions of the electromagnetic
plasma, the hydrodynamic approach is extremely useful in describing
many bulk and collective properties of the plasma.
In fact the equations of hydrodynamics have been successfully employed to study
variety of non-linear and non-equilibrium phenomena in electromagnetic
plasma \cite{qedplasma}. However, in the literature of quark-gluon plasma (QGP),
the application of hydrodynamics in studying the collective and bulk-properties
of the plasma have not been fully exploited \cite{MM06}. The application of 
hydrodynamics is generally restricted to investigate the evolution of thermalized 
QGP only. It should be noted here that there are certain issues related with
the application of the hydrodynamics to the thermalized QGP still remains \cite{thermal}
like inclusion of viscous effects, formation time etc.
On the other hand the transport theory is employed to investigate 
the properties of non-equilibrium QGP \cite{kinetic}. 
It must be emphasized that in the past, inspired by the success of
hydrodynamics for electromagnetic plasmas, the non-equilibreted QGP related issues
like two-stream instability and thermalization of the plasma were studied
using the chromo-hydrodynamics(CHD) equations \cite{bhatt94b}. 
The linear perturbative analysis 
done in Ref.\cite{bhatt94b} was shown to give the results similar to that obtained from 
the kinetic theory \cite{kinetic}. Moreover,
the hydrodynamics approach was also used to study various collective
phenomena in
QGP \cite{kaja80,holm,mrow88,bhatt89,bhatt94a,sudip,bistrovic,Manuel}.
Usually in hydrodynamics one introduces a length $L$  and a time $\tau$ 
which characterize distance and time over which the plasma quantities change
significantly. For the validity of hydrodynamics length associated
with the fluid element $\Delta V$, should be satisfying $(\Delta V)^{1/3}\ll L$ 
and $(\Delta V)^{1/3}\gg \lambda$, where $\lambda$ is the mean-free path. This
allows the fluid elements to persists over a several mean-free paths or for a several
collision frequencies. However in non-equilibrium plasmas for collective modes in collisionless limit 
another condition is more relevant:
In general each fluid element will have random velocity
$v_{thermal}$ and a flow velocity component $U$. If $U$ is same for all  particles
(if the fields acting on them are same) then the fluid element can
persist if  wave-vector $k$ and frequency $\omega$ satisfy 
the condition $k v_{thermal}\ll \omega$.
Thus in a few wave periods thermal effects do not spread the particles apart 
by as much as a wavelength and the fluid element essentially stays in tact , bound by collective fields.
This is quite unlike a neutral gas
where there is no long-range self-consistent field which can hold
the fluid elements in cold-collision less limit and therefore in that
case hydrodynamics might be meaningless (see for example Ref.\cite{tsy}).
Thus hydrodynamical equations has been applied to study various collective
phenomena in both QED and quark-gluon-plasma \cite{bhatt94a,MM06,qedplasma}.
However, hydrodynamics approach can be inadequate to describe certain
non-equilibrium phenomena where the momentum/velocity description of the
particle play an important role \cite{qedplasma}. At least in the linear
regime, for a quark-gluon plasma, it is found that the dispersion-relation
for the collective modes calculated either using kinetic-theory or
hydrodynamics remain same if the velocity dependent phenomena like the Landau
damping are ignored.

At present there are two versions of CHD equations in the literature of quark-gluon
plasma \cite{kaja80, MM06}. Therefore it is important, in our opinion,
to look for the the possible relationship between these two versions of the
hydrodynamics. This kind of study may be helpful in understanding the
assumptions under which the given set of CHD equations is valid. This may in turn
help in choosing an appropriate hydrodynamical equations for
a given physical problem. 

In what follows, we first briefly review
the two versions of CHD equations  given in Refs. \cite{kaja80,MM06}:
CHD equations were obtained from the QCD Lagrangian with some heuristic
arguments in Ref.\cite{kaja80}. Later more systemic derivation using
the kinetic-theory was given in Ref.\cite{hein}. However in Ref.\cite{hein}
the application to the pre-equilibrium phenomena was not discussed. 
These equations are give below:
\begin{eqnarray}
\partial_\mu\,N_\a^\mu &=&0,\\
\partial_\mu\,T_\a^{\mu\nu} &=& -gj_{\a\mu}^bF_b^{\mu\nu},\\
U^{\mu}_\a\partial_{\mu}I_{\a a}&=&-f_{abc}U^\mu_\a\,A_{\mu\,b}I_{\a c},
\label{kajaeq1}
\end{eqnarray}
\noindent
where, the index $\a$ denotes a stream or a specie (quark,anti-quark, gluons etc), 
$N_\a^\mu=\tilde{n}_\a\,U_\a^\mu$ is postulated to be the flux of specie $\a$ in a way similar
to \cite{MM06}. The four hydrodynamical velocity $U_\a^\mu$ is a colour singlet it
can be written in terms of three velocities $v_\a$ as $\gamma_\a\left(1,v_\a\right)$
with $\gamma_\a=1/\sqrt{1-v_\a^2}$. The quantity $Q_\a$ transforms covariantly under 
the gauge transformations and it can be regarded as a non-abelian "colour-charge" of
the fluid element. It is equation (3) that 
makes the CHD equations of Ref.\cite{kaja80} completely different from the equations
of electromagnetic plasmas.
Finally  the energy momentum tensor $T_\a^{\mu\nu}$ can be written as follows
\begin{equation}
T_\a^{\mu\nu}=\left(\tilde{\epsilon}_\a\,+\,\tilde{p}_\a\right)U_\a^\mu\,U_\a^\nu 
- \tilde{p}_\a\,g^{\mu\nu}
\label{emkaj}
\end{equation}
where $\tilde{\epsilon}_\a$ and $\tilde{p}_\a$ are energy density and pressure for 
the specie $a$. It must be noted that in this approach all the hydrodynamical
variables namely $\tilde{n}_\a, U_\a^\mu, \tilde{\epsilon}_\a, \tilde{p}_\a $
 are considered to be the colour scalars except $I_\a$. 
Equations \ref{kajaeq1} need to be supplemented by the equation of state and Yang-Mills
equations. The current density $j^\mu$ can be defined using the hydrodynamical variables
as follows
\begin{equation}
j_\a^\mu=g\Sigma_\a\,I_\a\,n_\a\,U_\a^\mu.
\end{equation}
 
The alternative formulation of the CHD equations \cite{MM06} regard the hydrodynamical
variables  as $N\times\,N$-matrices for $SU(N)$-colour group.
In this approach we have the following set of CHD equations:
\begin{eqnarray}
D_\mu n_\a^\mu &=&0,\\
D_\mu t_\a^{\mu\nu}&=&\frac{g}{2}\left\{F_\mu^\nu,n_\a^{\mu}\right\}
\label{mm06}
\end{eqnarray}
\noindent
where,$\a$ labels the stream or species and the flux $n_\a^\mu$ is defined 
as $n_\a^\mu=\,n_\a u_\a^\mu$. $D_\mu\,=\partial_\mu-ig\left[A_\mu,...\right]$ is the
covariant derivative and $\left\{...,...\right\}$ denote the anti-commutator.
The energy-momentum tensor $T_\a^{\mu\nu}$ for this version of CHD differs from
the equation (\ref{emkaj}) and it is defined as
\begin{equation}
t_\a^{\mu\nu}=\frac{1}{2}\left(\epsilon_\a\,+\,p_\a\right)\left\{u_\a^\mu\,,u_\a^\nu\right\} 
- p_\a\,g^{\mu\nu}.
\label{emm}
\end{equation}
The current density in the fundamental representation defined as 
\begin{equation}
j^\mu=-\frac{g}{2}\Sigma_\a\left(n_\a u_\a^\mu-\frac{1}{N}tr\left[n_\a u_\a^\mu\right]\right)
\end{equation}
\noindent All the hydrodynamical variables namely $n_\a, u_\a^\mu, \epsilon$ and
$p$, in this formulation, are $N\times N$ matrices for $SU(N)$-group and transform
covariantly under the gauge transformations. Derivation of equations (6-7) from
the kinetic theory with colour covariant  quark and gluon distribution functions
was given in Ref.\cite{MM06}.

Let us first note that irrespective of the version of the CHD equations used, 
the covariant continuity equation is always satisfied i.e.
\begin{equation}
 D_\mu j^\mu =0.
\end{equation}
\noindent
For simplicity let us consider the case for $SU(2)$
group and assume the 'cold' plasma $p_\a=0$ and $\epsilon=m_\a n_\a$ with 
$m_\a$ being the mass of the particle in specis $\a$ . 
In this case we
can write equation (\ref{emm}) as
\begin{equation}
 t_\a^{\mu\nu}=\frac{m_\a n_\a}{2}\left\{u_\a^\mu,u_\a^\nu\right\}
\end{equation}

 Next, consider
the following decomposition of the quantities $n_\a$ and $u_\a^\mu$ appearing
in equations (8-9),
\begin{eqnarray}
 n_\a &=& \left(I_\a+I_{\a 0}\right) n_{\a s},\\
u_\a^\mu &=&  \left(Q_\a+Q_{\a0}\right)u_{\a s}^\mu
\label{decomp}
\end{eqnarray}
\noindent
where, we have introduced the scalar quantities $n_{\a s}$ and $u_{\a s}^\mu$
for the gauge invariant number density and velocities for a given specie
or a stream respectively. Quantities $\left(I_\a+I_{\a 0}\right)$ and 
$\left(Q_\a+Q_{\a 0}\right)$
transforms gauge covariantly and they can be represented as
$\left(I_{\a a}T^a+I_{\a 0}\mathbb{I}\right)$ and $\left(Q_{\a a}T^a+Q_{\a 0}\mathbb{I}\right)$ 
respectively with $T^a$ being the group generators and 
$\mathbb{I}$ is an identity. It ought to be noted here that for gluon
sector we need the adjoint representation as the gluon distribution function
in the underlying kinetic theory description represented by $(N^2-1)\times (N^2-1)$
matrix. For the present purpose we restrict ourselves for the
fundamental representations necessary for describing the quark sector only.
Generalization for the gluonic sector can be straightforward \cite{MM06}.
The total current density can be constructed as
a sum of the current generated by each specie i.e. $j^\mu=\Sigma_\a j_\a^\mu$ where,
\begin{equation}
j_\a^\mu=-\frac{g}{2}\left(n_\a u_\a^\mu-\frac{1}{N}tr\left[n_\a u_\a^\mu\right]\right).
\end{equation}
From the above definitions one can always write,
\begin{equation}
 n_\a u_\a^\mu = \mathcal{R}_\a(IQ) n_{\a s}u_{\a s}^\mu.
\end{equation}
\noindent
where, 
\begin{equation}
\mathcal{R}_\a(IQ)=\left[\, I_\a Q_\a\,+\,I_\a Q_{\a 0}+I_{\a 0}Q_{\a}\,
+\,I_{\a 0}Q_{\a 0}\mathbb{I}\,\right].\nonumber
\end{equation}

Thus  $tr\left[n_\a u_a^\mu\right] = \left[tr\left(IQ\right)+
NI_{\a 0}Q_{\a 0}\right]n_{\a s}u_{\a s}^\mu$
and $tr\left(IQ\right) = NI_bQ^b$, where the summation convention over the colour
index $b$ is implied. Thus one can write expression for the colour density as
\begin{equation}
 j_\a^\mu = \frac{g}{2}\left[\,\,\overline{I_\a Q_\a}\,
+\,I_{\a 0}Q_\a\,+I_\a Q_{\a 0}\right]
n_{\a s}u_{\a s}^{\mu}
\end{equation}
\noindent
where, $\overline{I_\a Q_\a}=I_\a Q_\a-\frac{1}{N}tr\left(I_\a Q_\a\right)$ is
a trace-less quantity. It is easy to check that $tr\left(j_\a^\mu\right)=0$
as it should be the case. We have replaced of $n_\a$
and $u_{\a}^\mu$ with a new (and more) set of variables 
$\left (I_\a + I_{\a 0}\right)$, $\left (Q_\a + Q_{\a 0}\right)$
$n_{\a s}$ and $u_{\a s}^\mu$. Since we have introduced two more variables
than the originals, we have a freedom to impose two conditions on
them.
Before  exercising this freedom, let us first note that colour singlet
flux $n_{\a s} u_{\a s}^\mu$ may be a conserved quantity in absence of
any collisions i.e.,
\begin{equation}
 \partial_\mu \left(n_{\a s}u_{\a s}^\mu\right) = 0.
\end{equation}
\noindent
After substituting for   $j_\a^\mu$ into $D_\mu j_{\a}^\mu=0$ and after using 
equations (16-17) one gets,
 \begin{equation}
  u_{\a s}^\mu D_\mu \left[\overline{IQ}+I_\a Q_{\a 0} + I_{\a 0}Q_\a \right]=0
\label{sflux}
 \end{equation}
This equation has similar structure with the equation of colour charge dynamics
given by equation (3). 

Next consider the "Lorentz-force" term described by the right hand-side of equation 
(7), by adding and subtracting terms $\frac{1}{N}tr\left(IQ\right)$
in $n_\a u_{\a s}^\mu$ one can write
\begin{widetext}
\begin{equation}
 \left\{ F_{\mu}^\nu\,,\,n_{\a}u_\a^\mu \right\} = \left\{ T^d\, , \,
j_\a^\mu+\left(\frac{1}{N}tr\left(I_\a Q_\a\right)\mathbb{I}+
I_{\a 0}Q_{\a 0}\mathbb{I}\right) n_{\a s}u_{\a s}^\mu     \right\}F_{d \mu}^\nu
\end{equation}
\end{widetext}
\noindent
where, we have used $F_\mu^\nu=F_{d \mu}^\nu T^d$. From this one can
find 
\begin{equation}
 tr\left\{ F_{\mu}^\nu\,,\,n_{\a}u_\a^\mu \right\}=
2 j_{\a b}^\mu F_{b \mu}^\nu
\end{equation}
\noindent
where, we have used $T^d j_\a^\mu=\frac{1}{N}\delta^{db}j_{\a b}^\mu\mathbb{I}$.
Thus the trace of the "Lorentz-force" term may help us in identifying
$\left[\,\,\overline{I_\a Q_\a}\,
+\,I_{\a 0}Q_\a\,+I_\a Q_{\a 0}\right] $ as the colour charge similar
 \cite{kaja80} and it obeys the same differential equation [equation (3)].
 
Next, consider $t_\a^{\mu\nu}$ in terms of the new variables 
\begin{widetext}
\begin{equation}
 t_\a^{\mu\nu} = 2\left[\overline{I_\a Q_\a}+I_{\a }Q_{\a 0}+I_{\a 0}Q_\a
+\frac{1}{N}tr\left(I_\a Q_\a\right)+I_{\a 0}Q_{\a 0}\right]
\left(Q_\a+ Q_{\a 0}\right)n_{\a s}u_{\a s}^\mu u_{\a s}^\nu
\end{equation}
\end{widetext}
Next we can exercise our freedom and impose the following conditions: 
\begin{eqnarray}
 u_{\a s}^\mu D_\mu \left(I_\a\right) &=& 0\\
u_{\a s}^\mu D_\mu \left(Q_\a\right) &=&0
\end{eqnarray}
From this it is easy to see that $I_{\a a}^2$ and $Q_{\a a}^2$ are the
constants of motions and one may identify $I_{\a 0}=\sqrt{I_{\a a}^2}$
and $Q_{\a 0}= \sqrt{Q_{\a a}^2}$. It is worth noting
that dynamical equations for both $I$ and $Q$ are having the same
form as the colour-charge dynamics equation of Ref.\cite{kaja80}.
After the imposition of the above conditions, equation (18) can
be written as
\begin{equation}
 u_{\a s}^\mu D_\mu \left(\overline{IQ}\right)=0.
\end{equation}
 The covariant derivative of $t_\a^{\mu\nu}$ can be written as
\begin{widetext}
\begin{equation}
 D_\mu t_{\a}^{\mu\nu}=2\left[\overline{I_\a Q_\a}+I_{\a }Q_{\a 0}+I_{\a 0}Q_\a
+\frac{1}{N}tr\left(I_\a Q_\a\right)\mathbb{I}+I_{\a 0}Q_{\a 0}\mathbb{I}\right]
\left(Q_\a+ Q_{\a 0}\right)\partial_\mu\left(n_{\a s}u_{\a s}^\mu u_{\a s}^\nu  \right)
\end{equation}
\end{widetext}

By taking trace of $t_{\a }^{\mu\nu}$ expression and equating it with (20)
we get
\begin{equation}
 \mathcal{N}n_{\a s}u_{\a s}^{\mu}\partial_\mu u_{\a}^\mu=gj_{\a b}^\mu F_{b \mu}^\nu
\end{equation}
\noindent
where, $\mathcal{N}$ is a function of time in general and it is defined a trace of 
the following expression:
\begin{widetext}
 \begin{equation}
  \mathcal{N}=tr\left\{\left[\overline{I_\a Q_\a}+I_{\a }Q_{\a 0}+I_{\a 0}Q_\a
+\frac{1}{N}tr\left(I_\a Q_\a\right)\mathbb{I}+I_{\a 0}Q_{\a 0}\mathbb{I}\right]
\left(Q_\a+ Q_{\a 0}\right)\right\}.
 \end{equation}
\end{widetext}
It should be noted that $\mathcal{N}$ is a function
of space and time in general. But one can note that $N$ can
be calculated using dynamics of $I_\a$ and $Q_\a$  given by equations (22-23). 
Therefore if at some initial time $t_0$ if  $I_\a(t_0) =Q_\a(t_0)$

However it can be simplified
if one notes that the dynamics of $Q_\a$ and $I_\a$ are being described by the 
similar kind of equations [i.e. equations (22-23)]. Therefore if at some initial time
their values are taken to be same then for all the subsequent times
they will remain same i.e. $I_\a(t)=Q_\a(t)$. Under
these conditions $\mathcal{N} =4\left(Q_{\a a}^2\right)^{3/2}$ can become
a constant.
This would allow for the redefinition of
the colour charge in equations (18-21) by dividing it by the factor
$\left[\,\,\overline{I_\a Q_\a}\,+\,I_{\a 0}Q_\a\,+I_\a Q_{\a 0}\right]$ 
appearing in equations (18, 26)
by $4\left(Q_{\a a}^2\right)^{3/2}$. Thus we have shown that when the charge
associated with the charge density $\rho_\a$ i.e. $I_\a$ and that
with the matrix velocity $U^\mu_\a$ i.e. $Q_\a$ are set to be same
at some initial time, then the both hydrodynamics we discussed here
become equivalent. 

In conclusion, we have discussed the relationship between the two
versions of the equations of colour hydrodynamics. We have shown
that under the assumption that the charges associated with 
density and velocity variables of the matrix hydrodynamics are same
at some initial time then the both versions become identical. However,
in deriving we have ignored pressure gradients for the simplicity. 
But terms with the pressure gradients can be incorporated
by some straightforward arguments.

\noindent
{\bf Acknowledgements } 
 We would like thank St.  Mr\' owczy\' nski for useful discussions.

\end{document}